\documentclass[12pt]{article}
\usepackage{amsmath,amssymb}
\usepackage{eucal}

\usepackage[dviwindo]{graphicx}

\newcommand{\bs}{\begin{subequations}}
\newcommand{\es}{\end{subequations}}
\numberwithin{equation}{section}
\def \myfigures #1#2#3#4
{\begin{figure}[ht]
    \begin{center}
        \includegraphics[width=#2 \textwidth]{#1.eps}
        \hfill
        \includegraphics[width=#4 \textwidth]{#3.eps}
    \end{center}
\end{figure}}

\newcommand{\ben}{\begin{eqnarray}}
\newcommand{\een}{\end{eqnarray}}
\newcommand{\la}{\label}
\begin{document}

\title{On the Solutions of Einstein Equations with Massive Point Source}

\vskip 1.5truecm

\author{P.P. Fiziev\thanks{Department of Theoretical Physics, University of
Sofia,Boulevard 5 James Bourchier, Sofia 1164, Bulgaria
E-mail:\,\,\,fiziev@phys.uni-sofia.bg}}

\date{}
\maketitle

\begin{abstract}
We show that Einstein equations are compatible with the presence
of massive point particles and find corresponding two parameter
family of solutions. They are defined by the bare mechanical mass
$M_0>0$ and the Keplerian mass $M<M_0$ of the point source of
gravity. The global analytical properties of these solutions in
the complex plane define a unique preferable radial variable of
the problem.
\end{abstract}

\sloppy

\section{Introduction}

A clear physical motivation for consideration of massive point
particle sources of gravitational field in GR, both electrically
neutral and charged ones, can be found as early as in 1962-63
Feynman lectures on gravity \cite{Feynman}. In spite of this fact
the problem is still open.

At present the vast majority of relativists do not accept the
consideration of point particles in general relativity (GR) as
incompatible with Einstein equations (EE) idealization. There are
different reasons: some doubts about consistence of the theory of
mathematical distributions (like Dirac $\delta$-function
$\delta({\bf r})$) with the nonlinear character of EE; the clear
understanding of the drastic change of geometry of the Riemannian
space-time $\mathbb{M}^{(1,3)}\{g_{\mu\nu}\}$ in a vicinity of a
point with infinite concentration of energy in it, etc.

On the other hand it is obvious that in Nature very distant
objects like stars look like "points" of finite mass and finite
luminosity. This fact has a proper mathematical description in
Newton theory of gravity in the language of mathematical
distributions. A formal mathematical problem is to find a
corresponding idealized treatment of such objects in GR, as well,
but up to recently no reasonable approach was known.

Here we show that correct mathematical solutions of EE with
$\delta({\bf r})$ term in the rhs do exist. Such solutions
describe a two parameter family of analytical space-times
$\mathbb{M}^{(1,3)}\{g_{\mu\nu}\}$ with specific strong
singularity at the place of the massive point source with bare
mechanical mass $M_0>0$ and Keplerian mass $M<M_0$.

The price, one has to pay for this enlargement of standard GR
framework, is:

1) To consider metric coefficients $g_{\mu\nu}(x)$ as functions of
class ${\cal C}^0$ of the coordinates $x$. Some of them are to
have definite finite jumps in their first derivatives at the place
of the point source, necessary to reproduce the $\delta({\bf r})$
term in the lhs of EE from the Einstein tensor $G_\mu^\nu\sim
\delta({\bf r})$.

2) To accept the unusual geometry of space-time around the matter
point in GR which actually appeared at first in the original
Schwarzschild article \cite{Schwarzschild} and has been discussed
by Brillouin \cite{Brillouin} as early as in 1923. This geometry
is essentially different from the geometry around space-time
points with finite energy density in them. The global properties
of the space-time are different from the ones of popular
Hilbert-Droste-Weyl form of the Schwarzschild solution \cite{HDW}.

At present the original Schwarzschild geometry and other similar
geometries of space-time are widely ignored in GR. In addition,
the historical facts unfortunately are not reproduced in
completely correct way in the most of the modern literature on
this subject.

The important physical consequences, which can be derived using
the new solutions of EE, discussed in the present article, can be
considered after these solutions will be properly studied.

\section{Formulation of the Mathematical Problem}
In its proper frame the single point particle with bare mechanical
mass $M_0$, placed at the origin of the standard spherical
coordinate system in the 3D Riemannian space
$\mathbb{M}^{(3)}\{g_{ij}\}\subset\mathbb{M}^{(1,3)}\{g_{\mu\nu}\}$
yields the familiar static metric \cite{books}
$$ds^2\!=\!g_{tt}(r)\,dt^2\!+\!g_{rr}(r)\,dr^2\!-\!\rho(r)^2(d\theta^2\!+
\!\sin^2\theta\,d\phi^2)$$
with three unknown functions
$g_{tt}(r)\geq 0,\,g_{rr}(r)\leq 0,\,\rho(r)$ of the radial
variable $r\geq 0$. The variable $r$ is not defined by the $SO(3)$
symmetry of the problem, nor by its global-time translation
invariance with respect to the group $T_t(1)$. The only clear
thing is that the value $r=0$ must correspond to the center of
spherical symmetry, $C$, where the massive matter point is placed.

In contrast, the quantity $\rho$ has a clear geometrical meaning:
it defines the area $A_\rho=4\pi\rho^2$ of the sphere, centered at
the point $C$. Physically $\rho$ defines the luminosity of the
objects, which is reciprocal to $A_\rho$. Therefore we shall refer
to $\rho$ as "a luminosity variable".

At first glance the function $\rho(r)$ may be chosen in quite
arbitrary way, thus fixing the remaining (radial) gauge freedom of
the problem -- the only one, which is not fixed by symmetry
reasons. In this article we shall show that choosing a definite
class of functions $\rho(r)$ one can solve correctly the Einstein
equations:
\ben G^\mu_\nu = \kappa T^\mu_\nu \la{Einst}\een
with stress-energy tensor $T^\mu_\nu\sim M_0\delta({\bf r})$,
which describes a massive point source with bare mass $M_0$ and
mechanical action ${\cal A}_{M_{0}}\!=\!-M_0\int\!ds$.

It may seems strange that to solve the problem, one needs to fix
the class of coordinates by choosing the radial gauge. As we shall
see, the proper choice of coordinates is dictated naturally by the
boundary conditions. In the problem at hand these conditions are
masked in Dirac $\delta$-function, which describes in a formal
mathematical way the properties of source of gravity and its
boundary.

The shortest way to derive the field equation of our specific
problem is to perform a restriction on the orbits of the group
$SO(3)\times T_t(1)$ of the total action ${\cal A}_{tot}\!=\!{\cal
A}_{GR}+{\cal A}_{M_{0}}$, where ${\cal A}_{GR}\!=\!-{1\over{16\pi
G_n}}\int\!d^4x\sqrt{|g|}R$ is the Hilbert-Einstein action.
Neglecting the inessential surface terms and using $r$ as an
independent variable, we arrive at one dimensional problem with
Lagrangian
$${\cal L}={\frac {1} {2 G_N}}\!\left({
{2\rho\rho^\prime\left(\sqrt{g_{tt}}\right)^\prime\!+\!
\left(\rho^\prime\right)^2\sqrt{g_{tt}} } \over {\sqrt{\!-g_{rr}}}
}\!+\!\sqrt{g_{tt}}\sqrt{\!-g_{rr}}\right) -
M_0\sqrt{g_{tt}}\delta(r).$$
Here $G_N$ is the Newton
gravitational constant, $\delta(r)$ is the 1D Dirac
$\delta$-function \cite{Gelfand}. (We are using units $c\!=\!1$.)
The corresponding Euler-Lagrange equations read: \ben
\left({{2\rho\rho^\prime}\over{\sqrt{\!-g_{rr}}}}\right)^\prime
-{{{\rho^\prime}^2}\over{\sqrt{\!-g_{rr}}}}-\sqrt{\!-g_{rr}}+2 G_N
M_0 \delta(r)=0,
\nonumber\\
\left({{\left(\rho\sqrt{\!g_{tt}}\right)^\prime}\over{\sqrt{\!-g_{rr}}}}
\right)^\prime
-{{\rho^\prime\left(\sqrt{\!g_{tt}}\right)^\prime}\over{\sqrt{\!-g_{rr}}}}=0,
\nonumber\\
{{2\rho\rho^\prime\left(\sqrt{g_{tt}}\right)^\prime\!+\!
\left(\rho^\prime\right)^2\sqrt{g_{tt}} } \over {\sqrt{\!-g_{rr}}}
}\!-\!\sqrt{g_{tt}}\sqrt{\!-g_{rr}}\stackrel{w}{=}0 \la{EL}\een
where the symbol "\,$\stackrel{w}{=}$\," denotes a weak equality
in the sense of theory of constrained dynamical systems. (As a
result of the rho-gauge freedom the field variable
$\sqrt{\!-g_{rr}}$ is not a true dynamical variable but rather
plays the role of a Lagrange multiplier, which is needed in a
description of constrained dynamics.) Of course, the same
equations (\ref{EL}) can be derived directly from EE (\ref{Einst}).

Outside the point source of the gravitational field (i.e. for
$r>0$, where $ \delta(r)\equiv 0$) one obtains the standard
solution of this system \cite{ComJann}: \ben
g_{tt}(r)=1-{{\rho_G}/{\rho(r)}},\,\,\, g_{tt}(r)\, g_{rr}(r)=
-\left(\rho(r)^\prime\right)^2, \la{ss}\een where $\rho_G=2G_N M$
is a new integration constant -- the Schwarzschild radius, $M$ is
the gravitational (Keplerian) mass of the source, and $\rho(r)$ is
an arbitrary ${\cal C}^2$ function of $r$.

As usual, this solution describes the gravitational field outside
the point source (and outside any body with spherically symmetric
mass distribution). According to the Birkhoff theorem \cite{books}
it is unique (up to the choice of the radial variable $r$).
Nevertheless, being a pure vacuum solution of EE, it is not able
to describe the mass distribution $M_0\delta(r)$ in the very
{\em point} source, as well as the mass distribution in any other
matter source.

To see this in a more transparent way, let us consider the Hilbert
gauge (HG) of the radial variable: $r\equiv \rho$ \cite{HDW}, in
which the field equations (\ref{EL}) can be rewritten in the form:
\ben \left(\!\sqrt{\!-g_{\rho\rho}}\!-\!{1
\over{\sqrt{\!-g_{\rho\rho}}}}\!\right)\!{d\over{d
\ln\rho}}\!\ln\!\left(\! \rho\left(\!{1\over
{\!-g_{\rho\rho}}}\!-\!1\!\right)\!\right)\!=\nonumber\\ 2
\sigma_0 G_N M_0\,\delta(\rho\!-\!\rho_0),\nonumber\\ {{d^2\ln
g_{tt}}\over{(d \ln\rho})^2}\!+\!{1\over 2}\left(\!{{d\ln
g_{tt}}\over{d \ln\rho}}\!\right)^2\!\!+\!\left(\!1\!+\!{1\over
2}{{d\ln g_{tt}}\over{d \ln\rho}}\!\right){{d\ln
g_{\rho\rho}}\over{d \ln\rho}}\!=\!0, \nonumber\\ {{d\ln
g_{tt}}\over{d \ln\rho}}+g_{\rho\rho}+1\stackrel{w}{=}0.
\la{ELrho}\een Here we have utilized the properties of Dirac
$\delta$-function and the assumption that the metric coefficients
are usual ${\cal C}^2$ functions of $r$ (not distributions). Then
the only remnant of the function $\rho(r)$ in the system
Eq.(\ref{ELrho}) are the numbers $\rho_0=\rho(0)$ and
$\sigma_0=sign(\rho^\prime(0))$, which enter only the first
equation, related to the source of gravity.

One usually ignores the general case of an arbitrary value
$\rho_0\!\neq\!0$ accepting the value $\rho_0=0$, which seems to
be natural in HG. Indeed, if we consider the luminosity variable
as a measure of the {\em real} geometrical distance to the point
source of gravity in the 3D space (which is not the case), we have
to accept the value $\rho_0=0$ for the position of the point
source. Otherwise the $\delta$-function term in the Eq.
(\ref{ELrho}) will describe a {\em shell} with radius $\rho_0\neq
0$, instead of a {\em point} source. Such conclusion is based on
the above interpretation of the variable $\rho$.

Actually the point source has to be described using the function
$\delta(r)$ and is placed at the point $r=0$ {\em by definition}.
There is no reason to change this original position of the source,
or the interpretation of the variables in the problem at hand. To
what value of the luminosity variable $\rho_0=\rho(0)$ corresponds
the {\em real} position of the point source is not known {\em a
priori}. This depends strongly on the choice of the rho-gauge. One
can not exclude such a nonstandard behavior of a physically or
mathematically reasonable rho-gauge function $\rho(r)$ which leads
to some value $\rho_0 \neq 0$. Physically this means that instead
to infinity, the luminosity of the {\em point} source will go to a
{\em finite} value, when the distance to the source goes to zero
(see, for example, \cite{Schwarzschild}). This interesting
possibility appears in curved space-times due to their unusual
geometrical properties and is not supported by our Euclidean
experience. If one accepts the value $\rho_0=0$, one has to
recognize that the HG singularity at $\rho=0$ will be {\em
space-like,} not time-like, because for $\rho\in [0,\rho_G]$ the
variables $\rho$ and $t$ are changing their roles. This will be a
quite unusual and {\em non-physical} property for physical source
of a physical field of any kind.

The solution of the subsystem formed by the last two equations of
(\ref{ELrho}) is well known:
\ben g_{tt}(\rho)\!=\!1\!-\!{{\rho_G}/{\rho}},\,\,\,
g_{\rho\rho}(\rho)=-1/g_{tt}(\rho)\la{HGsolution}\een
-- the Hilbert form of the Schwarzschild solution. Note that in
this subsystem one of the equations is a field equation, but the
other one is a constraint. However, these functions do not solve
the first of the Eq.(\ref{ELrho}) for any value of $\rho_0$, if
$M_0\neq 0$. Indeed, for these functions the left hand side of the
first field equation equals identically zero and does not have a
$\delta$-function-type singularity, in contrast to the right hand
side. Hence, the first field equation remains unsolved by the
functions (\ref{HGsolution}). Thus we see that the assumption that
$g_{tt}(r)$, $g_{rr}(r)$ and $\rho(r)$ are usual ${\cal C}^2$
smooth functions, instead of distributions, yields a
contradiction, if $M_0\neq 0$. Indeed, in this way one is not able
to describe correctly the gravitational field of a massive point
source of gravity in GR.

This mathematical result is the real basis of the widespread
opinion, according to which it's impossible to describe a massive
point in GR. Actually, the right conclusion is that the class of
${\cal C}^2$ (and even the class of ${\cal C}^1$) metrics is not
proper for this purpose. To reach such goal, the first derivative
with respect to the variable $r$ of at least one of the metric
coefficients $g_{\mu\nu}$ must have a strictly definite jump. It
is needed to reproduce the Dirac $\delta$-function in the
energy-momentum tensor $T^\mu_\nu$ of the {\em massive} point
particle via the Einstein tensor $G^\mu_\nu$. Otherwise the EE
(\ref{Einst}) can not be fulfilled.

The above consideration illustrates one more important juncture.
It is obvious that physical results of any theory must not depend
on the choice of the variables and, in particular, these results
must be invariant under changes of coordinates. This requirement
is a basic principle in GR. It is fulfilled for any already {\em
fixed} mathematical problem.

Nevertheless, the change of the interpretation of the variables
may change the {\em formulation} of the mathematical problem and
thus, the physical results, because we are using the variables
according to their meaning. For example, if we are considering the
luminosity variable $\rho$ as a radial variable of the problem, it
seems natural to put the point source at the point $\rho=0$. In
general, we may obtain a physically different model, if we are
considering another variable $r$ as a radial one. In this case we
shall place the source at a different geometrical point $r=0$,
which now seems to be the natural position for the center $C$. The
relation between these two geometrical "points" and between the
corresponding physical models strongly depends on the choice of
the function $\rho(r)$, i.e. on the radial gauge. Thus, applying
the same physical requirements in different "natural" variables,
we arrive at different physical theories, because we are solving
EE under different boundary conditions, coded in corresponding
Dirac $\delta$-functions. One has to find a theoretical or an
experimental reasons to resolve this essential ambiguity.

\section{Field Variables and Radial Gauges, Suitable for a
         Correct Treatment of the Point Source of Gravity}
An obstacle for the description of the gravitational field of a
point source at the initial stage of development of GR was the
absence of an adequate mathematical formalism. Even after the
development of the correct theory of mathematical distributions
\cite{Gelfand} there still exist an opinion that this theory is
inapplicable to GR because of the nonlinear character of Einstein
equations (\ref{Einst}), see for example \cite{YB}. In a
recent article \cite{GT} the authors have considered singular
lines and surfaces, using mathematical distributions. They have
stressed, that "there is apparently no viable treatment of point
particles as concentrated sources in GR".

Here we propose a novel approach to this problem, based on a
specific choice of the field variables in the metric: \ben ds^2=
\!e^{2\varphi_1}dt^2\!\!-\!e^{\!-2\varphi_1\!+4\varphi_2-2\bar\varphi}dr^2\!\!-\!
\bar\rho^2e^{\!-2\varphi_1\!+2\varphi_2}(d\theta^2\!\!+
\!\sin^2\!\theta d\phi^2) \la{nc} \een where $\varphi_1(r)$,
$\varphi_2(r)$ and $\bar\varphi(r)$ are unknown functions of the
variable $r$ and $\bar\rho$ is a constant -- the unit for
luminosity variable $\rho=\bar\rho\, e^{-\varphi_1+\varphi_2}$.
The corresponding form of the Lagrangian: \ben {\cal
L}\!=\!\!{1\over{2 G_N}}\!
\Bigl(e^{\bar\varphi}\left(\!-(\bar\rho\varphi_1^\prime)^2\!+
\!(\bar\rho\varphi_2^\prime)^2\right)
\!+\!e^{\!-\bar\varphi}e^{2\varphi_2}\Bigr)- M_0\,
e^{\varphi_1}\delta(r) \la{L2}\een shows that the field variables
$\varphi_1(r)$, $\varphi_2(r)$ and $\bar\varphi(r)$ diagonalize
the kinetic-like part in ${\cal L}$. Hence, they play the role of
{\em a normal} fields' coordinates in the problem at hand. Now the
field equations read: \ben \bar\Delta_r \varphi_1(r)= {\frac
{G_NM_0}{\bar\rho^2}} e^{\varphi_1(r)-\bar\varphi(r)}
\delta(r),\,\,\,\,\,\bar\Delta_r \varphi_2(r)= {\frac
{1}{\bar\rho^2}}
e^{2\left(\varphi_2(r)-\bar\varphi(r)\right)}\hskip 1.1truecm
\la{FEq}\een where
$\bar\Delta_r\!=\!e^{\!-\bar\varphi}{d\over{dr}} \left(
e^{\bar\varphi} {d \over{dr}} \right)$ is related to the radial
part of the 3D-Laplacean. The variation of the total action with
respect to the auxiliary variable $\bar\varphi$ gives the
constraint: \ben
e^{\bar\varphi}\left(-(\bar\rho\varphi_1^\prime)^2\!+\!(\bar\rho\varphi_2^\prime)^2
\right)-e^{-\bar\varphi}e^{2\varphi_2}\stackrel{w}{=}0.
\la{constraint}\een The advantage of the above normal fields'
coordinates is that when expressed in them the field
equations (\ref{FEq}) are linear with respect to the derivatives
of the unknown functions $\varphi_{1,2}(r)$. This circumstance
legitimates the correct application of the mathematical theory of
distributions and makes our normal coordinates privileged field
variables.

The choice of the function $\bar\varphi(r)$ fixes the rho-gauge in
the normal coordinates. We have to choose this function in a way
that makes  the first of the equations (\ref{FEq}) mathematically
meaningful. Note that this inhomogeneous equation is quasi-linear
and has a correct mathematical meaning if, and only if, the
condition $|\varphi_1(0)-\bar\varphi(0)|<\infty$ is satisfied.

Let's consider once again the domain $r>0$. In this domain the
first of the equations (\ref{FEq}) gives $\varphi_1(r)=C_1\int
e^{-\bar\varphi(r)}dr+C_2$ with arbitrary constants $C_{1,2}$.
Suppose that the function $\bar\varphi(r)$ has an asymptotic
$\exp(-\bar\varphi(r))\sim k r^n$  in the limit $r\to +0$ (with
some arbitrary constants $k$ and $n$). Then one easily obtains
$\varphi_1(r)-\bar\varphi(r)\sim C_1 k r^{n+1}/(n+1)+
n\ln{r}+\ln{k}+C_2$ -- if $n\neq -1$, and
$\varphi_1(r)-\bar\varphi(r)\sim (C_1k-1)\ln{r} +\ln{k}+C_2$ --
for $n=-1$. Now we see that one can satisfy the condition
$\lim_{r\to 0}|\varphi_1(r)-\bar\varphi(r)|= constant <\infty$ for
arbitrary values of the constants $C_{1,2}$ if, and only if $n=0$.
This means that we must have $\bar\varphi(r)\sim k=const \neq
\pm\infty$ for $r\to 0$. We call such gauges {\em regular gauges}
for the problem at hand. Then $\varphi_1(0)=const\neq \pm\infty$.
Obviously, the simplest choice of a regular gauge is
$\bar\varphi(r)\equiv 0$. Further on we shall use this {\em basic
regular gauge} (BRG). Other regular gauges defer from it by a
regular rho-gauge transformation which describes a diffeomorphism
of the fixed by the BRG gauge manifold
$\mathbb{M}^{(3)}\{g_{mn}({\bf r })\}$. In terms of the metric
components the BRG fixing condition reads $\rho^4
g_{tt}+\bar\rho^4g_{rr}=0$.

Under this gauge the field equations (\ref{FEq}) acquire a
simple quasi-linear form: \ben \varphi_1^{\prime\prime}(r)=
{\frac {G_NM_0}{\bar\rho^2}}\, e^{\varphi_1(0)}\,
\delta(r),\,\,\,\, \varphi_2^{\prime\prime}(r)= {\frac
{1}{\bar\rho^2}}\,e^{2\varphi_2(r)}\hskip .2truecm \la{FEq0}\een
and constraint (\ref{constraint}) is: \ben
-(\bar\rho\varphi_1^\prime)^2\!+\!(\bar\rho\varphi_2^\prime)^2
-e^{2\varphi_2}\stackrel{w}{=}0. \la{constraint0}\een

Hence, the BRG gauge $\bar\varphi(r)\equiv 0$ has the unique
property to split the system of field equations (\ref{FEq}) and
the constraint (\ref{constraint}) into three independent relations
(\ref{FEq0}), (\ref{constraint0}).

\section{Solution of Einstein Equations with Massive Point Source}

\subsection{Solution in the Basic Regular Gauge}
The new form of Einstein field equations (\ref{FEq0}) of our
problem has a simple general solution in terms of mathematical
distributions (called sometimes "generalized
functions"\cite{Gelfand}): \ben \varphi_1(r)\!=\!{{G_N
M_0}\over{\bar\rho^2}}\,e^{\varphi_1(0)}\bigl(\!\Theta(r)\!-\!\Theta(0)\!\bigr)\,r
\!+\!\varphi_1^\prime(0)\,r\!+\!\varphi_1(0),\nonumber\\
\varphi_2(r)\!=\!-\ln\left({1\over{\sqrt{2\varepsilon_2}}}
\sinh\left(\sqrt{2\varepsilon_2}
{{r_\infty-r}\over{\bar\rho}}\right)\right). \hskip 1.2truecm
\la{Gsol}\een The first expression in Eq.(\ref{Gsol}) represents a
distribution $\varphi_1(r)$. In it $\Theta(r)$ is the Heaviside
step function. Here we use the {\em additional assumption}
$\Theta(0):=1$. It gives a specific regularization of the
products, degrees and functions of the distribution $\Theta(r)$
and makes them definite.

The second expression $\varphi_2(r)$ in Eq.(\ref{Gsol}) is a usual
function of the variable $r$. The symbol $r_\infty$ is used as an
abbreviation for the constant expression
$r_\infty=sign\left(\varphi_2^\prime(0)\right)\bar\rho\,
\sinh\left(\sqrt{2\varepsilon_2}e^{-\varphi_2(0)}\right)/\sqrt{2\varepsilon_2}$.
The constants  \ben \varepsilon_1\!=\!-{1\over
2}\bar\rho^2{\varphi_1^\prime(r)}^2\!+\!{{G_N
M_0}\over{\bar\rho^2}}\varphi_1^{\prime}(0)\,e^{\varphi_1(0)}
\bigl(\!\Theta(r)\!-\!\Theta(0)\!\bigr),\,\,\,\,\,\,
\varepsilon_2\!=\!{1\over
2}\left(\bar\rho^2{\varphi_2^\prime(r)}^2\!-\!e^{2\varphi_2(r)}\right)
\la{epsilon_12}\een are the values of the corresponding first
integrals (\ref{epsilon_12}) of the differential equations
(\ref{FEq0}) for a given solution (\ref{Gsol}).

Then for the regular solutions (\ref{Gsol}) the
condition(\ref{constraint0}) reads: \ben
\varepsilon_1+\varepsilon_2+{{G_N
M_0}\over{\bar\rho^2}}e^{\varphi_1(0)}
\bigl(\!\Theta(r)\!-\!\Theta(0)\!\bigr)
\stackrel{w}{=}0.\la{constraint_sol} \een

An unexpected property of this relation is that it cannot be
satisfied for any value of the variable $r\in (-\infty,\infty)$,
because $\varepsilon_{1,2}$ are constants. The constraint
(\ref{constraint_sol}) can be satisfied either on the interval
$r\in [0,\infty)$, or on the interval $r\in (-\infty,0)$. If, from
physical reasons we chose it to be valid at only one point $r^*\in
[0,\infty)$, this relation will be satisfied on the whole interval
$r\in [0,\infty)$ and this interval will be the physically
admissible real domain of the radial variable. Thus one can see
that our approach gives a unique possibility {\em to derive} the
admissible real domain of the variable $r$ from the dynamical
constraint (\ref{constraint0}), i.e., this dynamical constraint
yields a geometrical constraint on the values of the radial
variable. As a result, in the physical domain the values of the
first integrals (\ref{epsilon_12}) are related by the standard
equation
\ben\varepsilon_{tot}=\varepsilon_1+\varepsilon_2\stackrel{w}{=}0,
\la{epsilon}\een which reflects the fact that our variation
problem is invariant under local re-parameterization of the
independent variable $r$. At the end, as a direct consequence of
relation (\ref{epsilon}) one obtains the inequality
$\varepsilon_2=-\varepsilon_1>0$, because in the real physical
domain $r\in [0,\infty)$ we have $\varepsilon_1=-{1\over
2}\bar\rho^2{\varphi_1^\prime(r)}^2=const<0$.

For the function $\rho_{BRG}(r)\geq 0$, which corresponds to the
BRG, we obtain \ben\rho_{BRG}(r)=\rho_G
\left(1-\exp\Big({4{{r-r_{\infty}}\over{\rho_G}}}\Big)\right)^{-1}.
\la{reg_rho}\een

Now we can fix the arbitrary integration constants $\varphi_1(0)$,
$\varphi_1^\prime(0)$, $r_\infty$ and $\varepsilon_2$:

1) Imposing a number of additional requirements: i) to have an
asymptotically flat space-time, taking into account that in the
present variable $r$ we have $\rho(r)\to\infty$ when $r \to
r_\infty$; ii) to have a correct Keplerian mass M for $\rho \to
\infty$; iii) to have a consistence with the relation
$g_{tt}g_{rr}+{\rho^\prime}^2=0$;

2) Making use of suitable choice of the units for the luminosity
variable in the form $\bar\rho=G_NM=\rho_G/2$;

3) Introducing a gravitational mass defect for the point particle
in the following way:

Representing the bare mechanical mass $M_0$ of the point source in
the form $M_0=\int_0^{r_\infty}
M_0\,\delta(r)dr=4\pi\int_0^{r_\infty}
\sqrt{-g_{rr}(r)}\,\rho^2(r)\,\mu(r)dr$, one obtains for the mass
distribution of the point particle the expression
$\mu(r)=M_0\,\delta(r)/\left(4\pi\sqrt{-g_{rr}(r)}\,\rho^2(r)\right)=M_0\,\delta_g(r)$,
where
$\delta_g(r):=\delta(r)/\left(4\pi\sqrt{-g_{rr}(r)}\rho^2(r)\right)$
is the 1D {\em invariant} Dirac delta function. The Keplerian
gravitational mass $M$ can be calculated using the Tolman formula
\cite{books}: \ben M = 4\pi\int_0^{r_\infty}
\rho^\prime(r)\rho^2(r)\mu(r)dr
=M_0\sqrt{g_{tt}(0)}.\la{MKepler}\een Here we use the relation
$\rho^\prime = \sqrt{-g_{tt}\,g_{rr}}$. As a result we reach the
relations:
$g_{tt}(0)\!=\!e^{2\varphi_1(0)}\!=\!\exp\!\left(\!-2{{r_\infty}\over{G_N
M }}\!\right)\!=\!\left(\!{{M}\over{M_0}}\!\right)^2\!\leq\!1$ and
$r_\infty\!=\!G_N M \ln\left({{M_0}\over{M}}\right)\!\geq 0$.
(Note that due to our convention $\Theta(0):=1$ the component
$g_{tt}(r)$ is a continuous function in the interval $r\in
[0,\infty)$ and $g_{tt}(0)=g_{tt}(+0)$ is a well defined
quantity.)

The ratio $\varrho={{M}\over{M_0}}=\sqrt{g_{tt}(0)}\in [0,1]$
describes the gravitational mass defect of the point particle as a
second physical parameter in the problem. The Keplerian mass $M$
and the ratio $\varrho$ define completely the solutions
(\ref{Gsol}).

Then for the initial constants of the problem one obtains: \ben
\varphi_1(0)=\ln\varrho ,\,\,\,\varphi_2(0)=- \ln { {1-\varrho^2}
\over {2\varrho} } ,\nonumber\\ \,\nonumber\\ \varphi_1^\prime(0)=
{{1}\over{G_N M}}, \,\,\,\varphi_2^\prime(0)={{1}\over{G_N
M}}\,{{1+\varrho^2}\over{1-\varrho^2}}. \la{In_constMM}\een

Thus we arrive at the following form of the solutions
(\ref{Gsol}):

\ben \varphi_1(r)={{r\,\Theta(r)}\over{G_N M}}-\ln(1\!/\!\varrho),
\nonumber\\ \varphi_2(r)\!=\!-\ln\left( {1\over
2}\left({1\!/\!\varrho}\,e^{-r/G_N\!M}-
\varrho\,e^{r/G_N\!M}\right)\right) \la{sol_f} \een

and the rho-gauge fixing function
\ben\rho_{BRG}(r)=\rho_G
\left(1-\varrho^2\exp\Big({4{{r}/{\rho_G}}}\Big)\right)^{-1}.
\la{basic_rho_f}\een

An unexpected feature of this {\em two parametric} variety of
solutions for the gravitational field of a point particle is that
each solution must be considered only in the domain $r\in
\big[\,0,\,G_N M\ln\left({{1}/{\varrho}}\right)\big)$, if we wish
to have a monotonic increase of the luminosity variable in the
interval $[\rho_0,\infty)$.

It is easy to check that away from the source (i.e., for $r>0$)
the solutions (\ref{sol_f}) coincide with the HG solution and
acquire the well known standard form, when represented using the
variable $\rho$. This means that these solutions strictly respect
a generalized Birkhoff theorem. Its generalization requires only a
justification of the physical domain of variable $\rho$. In a
remarkable accord with Dirac's intuition \cite{Dirac} the minimal
value of the luminosity variable for the solutions (\ref{sol_f})
is \ben\rho_0=2G_N M /(1-\varrho^2)\geq \rho_G.\la{rho_0}\een This
changes the Gauss theorem and leads to a variety of different
important consequences. One of them is that we must apply the
Birkhoff theorem only in the interval
$\rho\in[\rho_0,\infty)\rightleftarrows r\in[0,r_\infty)$. As a
result, in this domain all {\em local} GR effects like
gravitational redshift, perihelion shift, deflection of light
rays, time-delay of signals, etc., will have their standard
{\em exact} values in gravitational field of the solutions
(\ref{sol_f}).

\subsection{Regular Mapping of the interval $r\in [0,r_\infty)$
to the whole interval $r\in[0,\infty)$}

It does not seem to be convenient to work with the unusual radial
variable $r\in [0,r_\infty)$. One can easily overcome this problem
using the regular radial gauge transformation \ben r \to
r_\infty{{r/\tilde r}\over{r/\tilde r+1 }}\la{rrgt}\een with an
arbitrary scale $\tilde r$ of the new radial variable $r$ (Note
that in the present article we are using the same notation $r$ for
different radial variables.) This linear fractional diffeomorfism
does not change the number and the character of the singular
points of the solutions in the whole compactified complex plane
$\widetilde{\mathbb{C}}_{\,r}$ of the variable $r$. The
transformation (\ref{rrgt}) simply places the point $r=r_\infty$
at infinity: $r=\infty$, at the same time preserving the
initial place of the origin $r=0$. Now the new variable $r$ varies
in the standard interval $r\in [0,\infty)$ and the regular solutions
(\ref{sol_f}) acquire the final form \ben
\varphi_1(r)=-\ln({1\!/\!\varrho}) \Bigg(1-{{r/\tilde
r}\over{r/\tilde r+1 }}\Theta\left({{r/\tilde
r}\over{r/\tilde r+1 }}\right)\Bigg),\nonumber\\
\varphi_2(r)=-\ln\left( {1\over 2
}\left((1\!/\!\varrho)^{{1}\over{r/\tilde
r+1}}-\varrho^{{1}\over{r/\tilde r+1}}\right)\right),\nonumber\\
\bar\varphi(r)=2\ln(r/\tilde r+1)+\ln(\tilde r/r_{\infty}).
\la{sol_NewGauge} \een The final form of the rho-gauge fixing
function reads: \ben\rho_{PRG}(r)=\rho_G
\left(1-\varrho^{{{2}\over{r/\tilde r +1 }}}\right)^{-1}.
\la{rho_NewGauge}\een

The last expression shows that the mathematically admissible
interval of values of the ratio $\varrho$ is the {\em open}
interval $(0,1)$. This is so, because for $\varrho=0$ and for
$\varrho=1$ we would have impermissible trivial gauge-fixing
functions $\rho_{PRG}(r)\!\equiv\!1$ and
$\rho_{PRG}(r)\!\equiv\!0$, respectively.

The expressions (\ref{sol_NewGauge}) and (\ref{rho_NewGauge})
still depend on the choice of unit for the new variable\, $r$. We
have to fix the arbitrary scale of this variable in the form
$\tilde
r=\rho_G/\ln(1/\varrho^2)=G_NM/\ln\left({{M_0}\over{M}}\right)$ to
ensure validity of the standard asymptotic expansion: $g_{tt}\sim
1 - \rho_G/r +{\cal O}\left((\rho_G/r)^2\right)$ when our {\em
last} radial variable $r$ goes to infinity. Then the final form of
the 4D interval, defined by the new regular solutions outside the
source (i.e., for $r>0$) is: \ben ds^2\!=\!e^{2\varphi_{\!{}_G}}
\left(dt^2\!-\!{{dr^2}\over{N_{\!{}_G}(r)^4}}\right)
\!-\!\rho_{PRG}(r)^2 \left(d\theta^2\!+\!\sin^2\!\theta
d\phi^2\right)\!.\hskip .2truecm \la{New_metric} \een Here we are
using a modified (Newton-like) gravitational potential: \ben
\varphi_{\!{}_G}(r;M,M_0):=-{{G_N M}\over {r+G_N M}/\ln({M_0\over
M})} \la{Gpot} \een a coefficient
$N_{\!{}_G}(r)=\left(2\varphi_{\!{}_G}\right)^{-1}
\left(e^{2\varphi_{\!{}_G}}-1\right)$, and a luminosity variable
\ben
\rho_{PRG}(r)\!=\!2G_NM/\left(1\!-\!e^{2\varphi_{\!{}_G}}\right)
\!=\!{{r\!+\!G_NM/\ln({M_0\over M})}\over {N_{\!{}_G}(r)}}.
\la{rhoG} \een

These basic formulas describe in a more usual way our regular
solutions of Einstein equations for  $r\in(0,\infty)$. Therefore
we call this radial gauge {\em physical radial gauge} (PRG).

The Eqs. (\ref{New_metric})-(\ref{rhoG}) show immediately that in
the limit $\varrho\to 0$ our solutions tend to the
Pugachev-Gun'ko-Menzel one \cite{PGM}, and
$\varphi_{\!{}_G}\left(0;M,M_0\right)=\ln\varrho \to -\infty$. In
the limit: $\varrho\to 1$ we obtain for any value of the ratio
$r/\rho_g$: $g_{tt}(r/\rho_G,\varrho)\to 1$,
$g_{rr}(r/\rho_G,\varrho)\to -1$, and $\rho(r/\rho_G,\varrho)\to
\infty$. Because of the last result the 4D geometry does not have
a  meaningful limit when $\varrho\to 1$. In this case
$\varphi_{\!{}_G}\left(r;M,M_0\right) \to 0$ at all 3D space
points. Physically this means that solutions with point source
without mass defect are not admissible in GR.

\section{Total energy of a point source and its gravitational
field}

In the problem at hand we have an extreme example of an "island
universe``. In it a privileged reference system and a well defined
global time exist. It is well known that under these conditions
the energy of the gravitational field can be defined unambiguously
\cite{books}. Moreover, we can calculate the total energy of the
aggregate of a mechanical particle and its gravitational field in
a canonical way. Indeed, the canonical procedure produces a total
Hamilton density ${\cal
H}_{tot}=\Sigma_{a=1,2;\mu=t,r}\,\pi_a^\mu\,\varphi_{a,\mu}-{\cal
L}_{tot}\!=\!{1\over{2G_N}}\left(-\bar\rho^2{\varphi_1^\prime}^2
+\bar\rho^2{\varphi_2^\prime}^2-e^{2\varphi_2}\right)+M_0
e^{\varphi_1}\delta(r)$. Using the constraint (\ref{epsilon}) and
the first of the relations (\ref{In_constMM}), one immediately
obtains for the total energy of the GR universe with one point
particle in it: \ben E_{tot}=\int_0^{{\infty}}{\cal H}_{tot}
dr=M=\varrho M_0< M_0\,.\la{E}\een

This result completely agrees with the strong equivalence
principle of GR. The energy of the gravitational field, created by
a point particle is the negative quantity:
$E_{GR}=E_{tot}-E_0=M-M_0=-M_0(1-\varrho)<0$.

The above consideration gives a clear physical explanation of the
gravitational mass defect for a point particle.

\section{Invariants of the Riemann Tensor}

To reach a coordinate independent description of the geometry of
space-time manifold with metric (\ref{nc}) we calculate the
invariants of the Riemann curvature tensor. Starting from the
general consideration in \cite{CMcL} we find in BRG
($\bar\varphi\equiv 0$) the following four independent invariants:
\ben
I_1:=e^{2(\varphi_1\!-\!2\varphi_2)}\varphi_1^{\prime\prime},\hskip
.truecm\nonumber\\
I_2:=e^{2(\varphi_1\!-\!2\varphi_2)}\left(\varphi_2^{\prime\prime}-
e^{2\varphi_2}/\bar\rho^2\right),\hskip .truecm\\
I_3:=e^{2(\varphi_1\!-\!2\varphi_2)} \left(-\varphi_1^\prime{}^2+
\varphi_2^\prime{}^2-e^{2\varphi_2}/\bar\rho^2
\right)/2,\nonumber\\
I_4:=e^{2(\varphi_1\!-\!2\varphi_2)}\varphi_1^{\prime}
\big(\varphi_1^{\prime}\!-\!\varphi_2^{\prime}\big)/2. \hskip
.truecm\nonumber \la{I}\een These are linear with respect to the
second derivatives of the functions $\varphi_{1,2}$ -- a property,
which is of critical importance when we have to work with
distributions $\varphi_{1,2}^{\prime\prime}$. For the regular
solutions (\ref{sol_f}) one obtains: \ben
I_1\!=\!{1\over{8\rho_G^2}}{{(1\!-\!\varrho^2)^4}\over{\varrho^2}}
\delta\left(\!{r\over{\rho_G}}\!\right)\!=\!-{1\over
2}R(r),\,\,\,\,\,
I_2=0,\hskip .truecm \nonumber\\
I_3\!=\!{{\Theta(r/\rho_G)\!-\!1}\over{8\rho_G^2\varrho^2}}\left(1\!-\!\varrho^2
e^{4r/\rho_G}\right)^4\!=
\!{{\rho_G^2}\over{8\varrho^2}}{{\Theta(r/\rho_G)\!-\!1}\over{\rho(r)^4}},\hskip
.truecm\nonumber\\
I_4\!=\!{{\Theta(r/\rho_G)}\over{4\rho_G^2}}\left(1\!-\!\varrho^2
e^{4r/\rho_G}\right)^3\!=
\!{{\rho_G}\over{4}}{{\Theta(r/\rho_G)}\over{\rho(r)^3}}. \hskip
.truecm \la{inv_sol}\een

The invariants $I_{1,...,4}$ of the Riemann tensor are well
defined distributions. This confirms the general expectations,
described in the articles \cite{TaubRaju}, where one can find a
correct mathematical treatment of distribution-valued curvature
tensors in GR.

As we see, the manifold $\mathbb{M}^{(1,3)}\{g_{\mu\nu}(x)\}$ for
our regular solutions has a definite geometrical singularity at
the point, where the physical source  -- the massive point
particle is placed. The fractional-linear transformation
(\ref{rrgt}) does not change the character of this singularity and
does not add new ones.

\section{Global Analytical Properties of the Solutions in
the Complex Plane of the Radial Variable}

To simplify the notations we introduce dimensionless variables:
$\zeta=r/G_NM$ -- instead of the radial variable $r$ in BRG,
$z=(r/G_NM)\ln(M_0/M)$ - instead of the radial variable $r$ in
PRG, and $w=\rho/\rho_G$ -- instead of the luminosity variable
$\rho$. In this Section we shall consider these variables as
complex valued ones in the corresponding complex plains
${\mathbb{C}}_{\,\zeta}$, ${\mathbb{C}}_{\,z}$,
${\mathbb{C}}_{\,w}$, and study the basic analytical properties of
the corresponding functions. Then, up to inessential constant
factors, the independent curvature invariants (\ref{inv_sol})
read: \ben I_1\!\sim\! \delta\left(\zeta\right),\,\,\,
I_3\!\sim\!\left(\Theta(\zeta)\!-\!1\right)\left(1\!-\!\varrho^2
e^{2\,\zeta}\right)^4,\,\,\,
I_4\!\sim\!\Theta(\zeta)\left(1\!-\!\varrho^2
e^{2\,\zeta}\right)^3 \,\,\,\hbox{ -- in BRG,
or}\la{Czeta_inv}\een \ben I_1\!\sim\!
\delta\!\left(\!z\!\right),\,\,
I_3\!\sim\!\left(\Theta\left({z\over{z\!+\!1}}\right)\!-\!1\right)
\left(1\!-\!\varrho^{2\over{z\!+\!1}}\right)^4,\,\,
I_4\!\sim\!\Theta\left({z\over{z\!+\!1}}\right)
\left(1\!-\!\varrho^{2\over{z\!+\!1}}\right)^3\hbox{ -- in
PRG.}\la{Cz_inv}\een As we see, the BRG-invariants  have a unique
analytical property: their {\em only} singularity in
${\mathbb{C}}_{\,\zeta}$ is the center $C$: $\zeta=0$. On the
compactified complex plain $\widetilde{\mathbb{C}}_{\,\zeta}$
these invariants have an additional essentially singular point
$\zeta=\infty$, where the jump of the luminosity variable
$w(\zeta)=\left(1\!-\!\varrho^2e^{2\,\zeta}\right)^{-1}$ on the
real axes equals 1. As seen from Eq. (\ref{Cz_inv}), the
transition to PRG, defined by the fractional linear mapping
(\ref{rrgt}), which now reads\ben
\zeta(z)=\big(\ln{1/\varrho}\big){z\over{z+1}}\la{zeta_z}
\,\,\,\rightleftharpoons\,\,\,
z(\zeta)={\zeta\over{\big(\ln{1/\varrho}\big)-\zeta}}\een simply
translates this essentially singular point at the position $z=-1$.
From analytical point of view the existence of such essentially
singular point is the basic difference between the descriptions of
massive point source in GR and in Newton gravity.

It is obvious that transition to any other radial variable z,
defined by nonlinear transformation $z=z(\zeta)$, more general
than a fractional linear one:
$z(\zeta)={{a+b\zeta}\over{c+d\zeta}}$, will create new singular
points in the plane $\widetilde{\mathbb{C}}_{\,z}$, or will change
the character of the existing ones, thus changing the mathematical
and the physical properties of the solution in the
$\widetilde{\mathbb{C}}_{\,z}$ plane.

Hence, from analytical point of view the PRG is a unique radial
gauge which defines {\em a preferable} radial variable $r$ in the
problem at hand, satisfying simultaneously the following two
conditions:

i) In the physical domain it varies in the natural interval $r\in
[0,\infty)$ and has no other singularities in this interval than
the very point source.

ii) The only singular points of the solution in the compactified
complex plain $\widetilde{\mathbb{C}}_r$ are: the place of the
source at $r=0$ and the unavoidable in GR essentially singular
point at $r=-1$, which is placed in the nonphysical domain.

This important result solves the longstanding problem of the
choice of radial variable $r$ for point source in GR on a clear
theoretical basis.

The only independent curvature invariant for the HG form of
Schwarzschild solution (\ref{HGsolution}) is
$\left(I_4\right)_{HG}\sim
w^{-3}=\left(1\!-\!\varrho^{2\over{z\!+\!1}}\right)^3$
\cite{books}. The comparison with Eqs. (\ref{Cz_inv}) makes it
clear that from geometrical point of view the HG solution of EE
(\ref{HGsolution}) is a vacuum one and essentially differs from
the massive point particle solutions.

Now we are ready to describe the singular character of the
coordinate transition from the Hilbert form of Schwarzschild
solution to the regular one (\ref{sol_NewGauge}), considered
outside the source. Eq. (\ref{rho_NewGauge}) shows that in
this domain the change of the coordinates is described, in both
directions, by the functions: \ben
w(z)=\left(1-\varrho^{2\over{z+1}}\right)^{-1}\,\,\,\rightleftharpoons
\,\,\, z(w)={{\ln(1/\varrho^2)}\over {\ln w - \ln(w-1)}}-1,
\,\,\,\varrho\in (0,1). \la{change} \een

The function $w(z)$ is regular at the place of the
point source $z=0$; it has a simple pole at $z=\infty$ and an
essentially singular point at $z=-1$. At the same time the inverse
function $z(w)$ has a logarithmic branch points both at the HG center
of symmetry  $w=0$ and the event horizon $w=1$. Thus we see how
one produces the HG singularities at $\rho=0$ and at
$\rho=\rho_{{}_H}$, starting from a regular solution. The
derivative
$$dz/dw={{\ln(1/\varrho^2)}\over{w(w-1)\left(\ln w-\ln(w-1)\right)^2}}$$
approaches infinity at these two points, hence the singular
character of the change of the variables in the whole complex
domain. The restriction of the change of the radial variables on
the corresponding {\em physical} interval outside the source: $z
\in (0, \infty) \rightleftharpoons w\in (1/(1-\varrho^2),
\infty)$, is a regular one.

\section{Conclusion}

In the present article we have studied a new, two parameter class
of solutions of Einstein equations. These static spherically
symmetric solutions describe the gravitational field of massive
point particle with bare mass $M_0>0$ and Keplerian mass $M$ ($0<M
< M_0$). The difference between these masses, or their ratio
$\varrho=M/M_0\in (0,1)$, define the gravitational mass defect for
the point particle. Such mass defect was not considered and
studied until now, because for the standard Hilbert form
(\ref{HGsolution}) of the Schwarzschild solution "the bare
rest-mass density is never even introduced" \cite{ADM} correctly.

The new solutions form a two parameter family of metrics on
singular manifolds $\mathbb{M}^{(1,3)}\{g_{\mu\nu}\}$, described
in details in the present article.

We have shown the principal role of the massive point source of
gravity. It presents a natural cutting factor for the {\em
physical} values of the luminosity variable $\rho\in [\rho_0,
\infty)$, where $\rho_0> \rho_G$ (\ref{rho_0}). This happens
because the infinite mass density of the matter point changes
drastically the geometry of the space-time around it.

Similar geometry of space-time with $\rho_0 \equiv \rho_G$ was
discovered at first in the original article by Schwarzschild
\cite{Schwarzschild}. According to Eq. (\ref{rho_0}), such
limiting value of the luminosity variable corresponds to zero
value $\varrho=0$ of mass defect ratio. For finite value of $M$
this is possible only if $M_0=\infty$. In this sense our work is a
proper extension of the Schwarzschild one to the physically and
mathematically admissible values of the mass defect ratio
$\varrho\in (0,1)$.

In full accord with Dirac's suggestion \cite{Dirac} our cutting of
the domain of luminosity variable places the event horizon in the
nonphysical domain of the variables. This effect is well known
from the solutions of Einstein equations with massive matter
sources of finite dimension.

The mathematical and the physical properties of the new solutions
are essentially different in comparison with the well known other
spherically symmetric static solutions to the Einstein equations.
All they have different type of singularities at the center of the
symmetry, which is surrounded by empty space. The previously known
solutions were often erroneously considered as a solutions for
describing of single point mass in GR.

It is clear that our solutions in generalized functions define in
mathematical sense the fundamental solutions of Einstein
equations, which are complete analogous to the fundamental
solutions of Poisson equation in Newton theory of gravity. Thus
the problem, formulated by Feynman in \cite{Feynman} is solved.

Further study of the new solutions and their physical applications
will be given in separate articles.

\vskip 1truecm

{\em \bf Acknowledgments}

\vskip .3truecm

The author is grateful to the High Energy Physics Division, ICTP,
Trieste, for the hospitality and for the nice working conditions
during his visit in the autumn of 2003 where an essential
development of the present article was done. He is grateful, too
to the JINR, Dubna, for the financial support of the present
article and for the hospitality and good working conditions during
his three months visits in 2003 and in 2004.

\end{document}